\documentclass[usenatbib]{mnras}
\usepackage{graphicx}	
\usepackage{amsmath}	
\usepackage{amssymb}	
\usepackage{mathptmx}
\usepackage{txfonts}
\usepackage[T1]{fontenc}

\title{On the Formation of Eccentric Millisecond Pulsars by Accretion-induced Collapse of Massive White Dwarfs}
\author[D. Wang and B.P. Gong]{
D. Wang, $^{1}$\thanks{E-mail: wangdi17@hust.edu.cn}
B.P. Gong, $^{1}$
\\
$^{1}$Department of Physics, Huazhong University of Science and Technology, Wuhan 430074, China
}

\date{Accepted 2023 September 25. Received 2023 September 25; in original form 2023 January 18}

\begin{document}
\label{firstpage}
\maketitle
\begin{abstract}
    
The millisecond pulsar(MSP) is believed to be an old neutron star(NS) having undergone spin-up by the accreting material from the donor. Whereas, the discovery of eccentric millisecond pulsars (eMSPs) in the Galactic field challenges such a scenario producing MSP-white dwarf (WD) only in the circular orbit. As orbital periods and companion mass of these eMSPs are located in a narrow range, a reasonable postulation is that they have the same origin. Although many models have been proposed to interpret their origin, however, the origin of the narrow range of the orbital period is still an open question. The accretion-induced collapse(AIC) of the ONe WD is considered to be an important pathway to form MSP, which was expected to result in the formation of MSP in the circular orbit due to tidal circularization. Here we revisited this scenario by the binary population synthesis including the specific circularization calculation. Our results indicate that binaries with insufficient circularization in this scenario can evolve into the eMSPs. The narrow initial binary parameters required by insufficient circularization can naturally account for the narrow range of the orbital period. Although the evolution of WD's AIC process has not been well understood, the characteristic of a narrow range in the orbital period of eMSPs can still set constraints on the physics of their evolution.

\end{abstract}
\section{Introduction}
Millisecond pulsars(MSPs) are the old neutron stars(NSs) having gained angular momentum by accretion of material from their donor stars, called the recycling model which is supported by many observations\citep{lorimer2008binary}. During the accreting stages, the orbit is expected to be circularized by the tidal interaction. The accretion is terminated while the donor stars evolve into the white dwarfs(WDs) so that the remnant MSP-WD systems are expected to be circular. 

However, the discovery of eccentric millisecond pulsars(eMSPs) in the Galactic field challenges such a recycling model. The first reported eMSP is PSR J1903+0327, which has a main-sequence star(MS) companion and a very large eccentricity $e \simeq 0.44$\citep{champion2008eccentric}. It is the only eMSP that has an MS companion, suggested to be formed from hierarchical triple system\citep{freire2011nature}. Subsequently, five eMSPs with WD companion are discovered\citep{bailes2010curious,barr2013northern,deneva2013goals,knispel2015einstein,camilo2015parkes}. Those binaries have similar narrow orbital period(22-32 d) and eccentricity(0.08-0.14). Their companions satisfy the mass-orbital period relation similar to the conventional pulsars\citep{ginzburg2022eccentric}, implying that they have undergone the same accretion process in the recycling process despite their much larger eccentricity. Such similarities suggest that they are not formed from the chaotic triple systems, producing systems with a  wider range of orbital periods. 

As a result, several models have been proposed to explain eMSP systems. \cite{antoniadis2014formation} suggested that the ejected material of the donor can form a circumbinary disk during the hydrogen flash process, then the dynamical interaction between such a disk and binary can excite a high eccentricity. In this model, the orbital period and eccentricity of the binary depend on properties of the disk such as lifetime, mass, position, and so on, which are very poorly understood. Consistent with the observation requires strong assumptions on the properties of the disk. \cite{han2021asymmetrical} and \cite{tang2023impact} proposed that the mass ejection from proto-WD could be asymmetrical, which thereby causes eccentricity via kick velocity regardless of whether the circumbinary disk is formed or not. Recently, \cite{ginzburg2022eccentric} found that the orbital periods of eMSPs are close to the convective eddy turnover time of the red giant. They thus proposed that the eccentricities of eMSPs are induced by the resonant convection.

Other models drop the assumption that neutron stars arise from the iron-core collapse of massive stars, as do conventional models. \cite{freire2013direct} suggested that the eMSP might be formed by the accretion-induced collapse(AIC) of a massive white dwarf. To avoid orbital circularization after the AIC from happening, the AIC event is required to take place while the companion has exhausted its envelope. They assumed that the WD is super-Chandrasekhar with high spin, which collapses while the critical mass reaches the current mass as spin-down. Then the orbital eccentricity is induced by the mass loss and kick via the AIC. Such a scenario is the so-called rotationally delayed accretion-induced collapse(RD-AIC) model which is then adopted by \citep{jiang2015strange,jiang2021strange}. They suggested that the eMSP is possibly produced by the RD-AIC onto the NS, resulting in the strange quark star(SS). The mass of the newborn neutron star is less than that of its progenitor because of the mass loss during the AIC process. Consequently,  the measured pulsar mass of two eMSPs being larger than the Chandrasekhar limit is difficult to be explained by the RD-AIC of the WD\citep{zhu2019mass,barr2017massive}. Though such RD-AIC is difficult to account for the narrow orbital period range of eMSP, it can easily interpret the presence of both high and low eccentric systems within this range, i.e. via stemming from different evolutionary channels.

The AIC evolution of WD has been extensively studied\citep{wang2018single,liu2018evolving,wang2022formation,tauris2013evolution}. When the mass of an accreting One WD reaches a critical mass, it collapses into NS. The eccentric orbit originating in the AIC is usually expected to be circularized during the subsequent accretion process. However, if the lifetime of the giant star companion after the onset of AIC is less than the circularization timescale, the NS-WD system left behind will be eccentric. Such insufficient circularized systems may exist in a small parameter space, which can explain the narrow orbital range observed. 

\cite{serylak2022eccentric} recently measured the companion mass of PSR J0955-6150 and found it to be significantly lower than predicted, a result they believe rules out all current models. Such deviation can be explained by the kick event that changes the orbital period. However, their calculation show that the RD-AIC model as a pure kick model always predicts larger eccentricities when the orbital period is met in accordance with observations. If the binary undergoes an insufficient circularization after the kick, then it can have a lower eccentricity than predicted by the RD-AIC model and thus satisfy the observation.

To investigate such a possibility, the orbital eccentric evolution needs to be considered in the AIC evolution of the WD. \cite{hurley2010formation} made a binary population synthesis(BPS) that contains orbital eccentric evolution. However, the distribution of eccentricity was not addressed in the literature. \cite{chen2011can} adopted the same BPS code to study the eccentric MSP with only the non-degenerate companion. 

In this paper, we make a similar BPS to study the eccentric MSP-WD systems and then compare them with the observations. In Section \ref{method}, the numerical method is exhibited. The simulation results are shown in Section \ref{result}. The discussion and summary are presented in Section \ref{discussion} and  Section \ref{summary} respectively.


\section{Numerical Methods}
\label{method}
Our simulation of the binary evolution is performed based on the rapid binary star evolution(BSE) code \citep{hurley2002evolution}. It uses the approximately analytic formula to describe the evolution of single stellar \citep{hurley2000comprehensive}. The orbital evolution includes the mass variation via Roche lobe overflow(RLO) and stellar winds, kick velocity of collapse and common envelope(CE) evolution, and the orbital angular momentum evolution via the gravitational wave radiation, magnetic braking, and tide.

Once the mass of the WD reaches $1.44 \mathrm{M}_{\sun}$ with mass growth by accretion, the oxygen-neon(ONe) WD will collapse into a NS of $1.3 \mathrm{M}_{\sun}$ while the carbon-oxygen(CO) WD will undergo a Type Ia supernova(SNe Ia). The eccentricity will be induced by the mass loss and kick velocity during the AIC \citep{hurley2002evolution}.

\subsection{The WD mass accumulation}
In the BSE, a white dwarf ejects some accreted material at low mass transfer rates and regains its envelope at high accretion rates to become a giant star\citep{hurley2002evolution,hurley2010formation}. It is different from the white dwarf mass growth model widely used in studies of accreting white dwarfs \citep{tauris2013evolution,wang2009helium,wang2014evolution,wang2018single}. We use this more commonly used model described below in this subsection and replace the original model in the BSE with it.

Following \cite{hachisu1999new}, the mass growth of the He-layer by the H-shell burning is dependent on the mass loss of the companion $\dot{M}_2$, given by 
\begin{equation}
    \dot{M}_{\mathrm{He}}=\eta_{\mathrm{H}}\left|\dot{M}_2\right|,
\end{equation}
where the mass accumulation efficiency of H-shell burning $\eta_{\mathrm{H}}$ is
\begin{equation}
    \eta_{\mathrm{H}}= \begin{cases}\dot{M}_{\mathrm{cr}} /\left|\dot{M}_2\right|, & \left|\dot{M}_2\right|>\dot{M}_{\mathrm{cr,H}} \\ 1, & \dot{M}_{\mathrm{cr,H}} \geq\left|\dot{M}_2\right| \geq \dot{M}_{\text {low,H }} \\ 0, & \left|\dot{M}_2\right|<\dot{M}_{\text {low,H }}\end{cases}.
    \end{equation}
Here the critical mass transfer rate for the steady hydrogen burning $\dot{M}_{\mathrm{cr,H}}$ is 
\begin{equation}
    \dot{M}_{\mathrm{cr,H}}=5.3 \times 10^{-7} \frac{(1.7-X)}{\mathrm{X}}\left(\frac{M_{\mathrm{WD}}}{\mathrm{M}_{\sun}}-0.4\right) \mathrm{M}_{\sun} \mathrm{yr}^{-1},
\end{equation}
where the $M_{\mathrm{WD}}$ is the WD mass and $X$ is the hydrogen mass fraction of the accreted material, and the minimum mass transfer rate at which a WD can accumulate mass by H-shell burning is $\dot{M}_{\text {low,H }}=\frac{1}{8}\dot{M}_{\mathrm{cr,H}}$.

The newborn He-layer will continue to burn, then the mass growth of the WD by the He-layer burning is given by
\begin{equation}
    \dot{M}_{\mathrm{WD}}=\eta_{\mathrm{He}} \cdot \dot{M}_{\mathrm{He}},
\end{equation}
where the mass accumulation efficiency of the He-layer burning $\eta_{\mathrm{He}}$ is
\begin{equation}
    \eta_{\mathrm{He}}= \begin{cases}\frac{\dot{M}_{\mathrm{cr,He}}}{\left|\dot{M}_{\mathrm{He}}\right|}, & \left|\dot{M}_{\mathrm{He}}\right|>\dot{M}_{\mathrm{cr,He}} \\\eta_{\mathrm{He}}^{\prime}, & \dot{M}_{\mathrm{cr,He}}>\left|\dot{M}_{\mathrm{He}}\right| \geq \dot{M}_{\mathrm{low,He}} \\ 0, & \left|\dot{M}_{\mathrm{He}}\right|<\dot{M}_{\mathrm{low,He}}\end{cases}.
    \end{equation}
The critical mass transfer rate for the steady He-layer burning is\citep{nomoto1982accreting}
\begin{equation}
    \dot{M}_{\mathrm{cr,He}}=7.2 \times 10^{-6}\left(\frac{M_{\mathrm{WD}}}{\mathrm{M}_{\sun}}-0.6\right) \mathrm{M}_{\sun} \mathrm{yr}^{-1},
    \end{equation}
and the minimum mass transfer rate of He-layer burning is\citep{woosley1986models}
\begin{equation}
    \dot{M}_{\text {low,He}}=4.0 \times 10^{-8} \mathrm{M}_{\sun} \mathrm{yr}^{-1}.
\end{equation}
And the $\eta_{\mathrm{He}}^{\prime}$ is taken from the linear interpolation of the numerical results by \cite{kato2004mass}.

The materials that can not be accumulated onto the WD's surface are assumed to be blown away as the wind carrying the specific orbital angular momentum of the WD \citep{hurley2002evolution}.

\subsection{Circularization}
If the circularization is insufficient after the occurrence of  AIC, then the remnant  NS-WD system will be eccentric. Thus the only important issue is the circularization mechanism of the NS and donor.

The circularization mechanism during stable mass transfer is tidal dissipation. In the tidal model used by the BSE, the tidal dissipation occurs in the envelope of the star, then the circularization efficiency is positively related to the envelope mass. The detailed formula can be found in section 2.3 of the \cite{hurley2002evolution}.

If the mass transfer between the NS and donor is unstable, the binary will enter the CE stage, after which the binary is usually believed to be circular. However, the simulation of the CE evolution of the eccentric binary shows that eccentric binaries are only partially circularized\citep{glanz2021common}. In the BSE, the CE evolution is described by the $\alpha\lambda$ model, in which the initial binding energy of the envelope is calculated by
\begin{equation}
    E_{\mathrm{bind}, \mathrm{i}}=-\frac{G}{\lambda}\left(\frac{M_1 M_{\mathrm{env} 1}}{R_1}+\frac{M_2 M_{\mathrm{env} 2}^{\prime}}{R_2}\right),
\end{equation}
and the initial orbital energy of the cores is 
\begin{equation}
    E_{\mathrm{orb}, \mathrm{i}}=-\frac{1}{2} \frac{G M_{\mathrm{c} 1} M_{\mathrm{c} 2}^{\prime}}{\mathrm{a}_{\mathrm{i}}}.
\end{equation}
Then orbital energy of the cores is transferred to the envelope during the CE evolution with an efficiency $\alpha$.\footnote{In the BSE, The defined initial orbital energy contains only the core component, which is not the same as the calculation using the entire star. The $\alpha\approx 3$ in the BSE corresponds to $\alpha=1$ in the latter \citep{hurley2010formation}, representing that all the orbital energy is transferred to the envelope.}

The CE stage is terminated while the envelope has enough energy to be ejected, so that the final orbital energy of the cores $E_{\mathrm{orb}, \mathrm{f}}$ is read, 
\begin{equation}
    E_{\mathrm{bind}, \mathrm{i}}=\alpha\left(E_{\mathrm{orb}, \mathrm{f}}-E_{\mathrm{orb}, \mathrm{i}}\right),
\end{equation}
where the final semi-axis $\mathrm{a}_{\mathrm{f}}$ can be derived by 

\begin{equation}
    E_{\mathrm{orb}, \mathrm{f}}=-\frac{1}{2} \frac{G M_{\mathrm{c} 1} M_{\mathrm{c} 2}^{\prime}}{\mathrm{a}_{\mathrm{f}}}.
\end{equation}

The initial circularized orbital energy of the cores is defined as $E_{\mathrm{circ}}=E_{\mathrm{orb}, \mathrm{i}}/(1-e^2)$. If $E_{\mathrm{orb}, \mathrm{f}}< E_{\mathrm{circ}}$, the final eccentricity is given by
\begin{equation}
    e_{f}=\sqrt{1-\frac{E_{\mathrm{orb}, \mathrm{f}}}{E_{\mathrm{circ}}}},
\end{equation}
and $e_f=0$ if $E_{\mathrm{orb}, \mathrm{f}}\ge E_{\mathrm{circ}}$.

\subsection{Criteria of the CE}
The criteria for entering the CE stage are also very critical. Whether or not a binary enters the CE phase after the AIC determines which circularization mechanism will work. Also, if the CE proceeds during the accretion phase of the WD, the short lifetime of the CE will prevent the WD from accreting enough mass for AIC.

Now the CE criteria is unclear. \cite{tauris2013evolution} adopted  $\dot{M}_2=3\dot{M}_{\mathrm{cr,H}}$ and \cite{wang2018single} used an unstable increases of $\dot{M}_2$ as the CE criteria. Thus they predict different  parameters space on the occurrence of AIC. The BSE adopts the  mass ratio as the CE criteria, which is not going to be changed by this work. But it is worth noting that the simulation results are strongly dependent on this assumption.

\subsection{Binary population synthesis}
On the choice of initial parameter of binaries, we adopt the setting of \cite{hurley2002evolution}, in which  all binaries are assumed to be initially circular. The initial parameter space of binaries is as follows: $0.8-80 \mathrm{M}_{\sun}$ for the primary mass $M_1$, $0.1-80 \mathrm{M}_{\sun}$ for the secondary mass $M_2$, $3-10^4 \mathrm{R}_{\sun}$ for the binary separation $a$. All parameters are taken from  $n_\chi$ grid points in the logarithmic space. Thus, the gird point interval is
\begin{equation}
    \delta \ln \chi = \frac{\ln \chi_{\max }-\ln \chi_{\min }}{n_\chi-1}
\end{equation}
for each initial binary parameter $\chi$($M_1$, $M_2$, $a$). The grid number $n_\chi$ we use is 400.

There are three evolutionary channels that depend on the type of companion star at the time of the AIC\citep{wang2020formation}: the MS channel, the red gaint(RG) channel, and the helium star(He) channel. The He channel will form more massive companions than observation \citep{tauris2013evolution}, which implies that eMSPs are formed from the MS/RG channel. Thus we only concern the detached eccentric($e > 0.01$) NS-WD systems formed from the AIC of WD in these channels. 

As long as a binary  evolves to such a system, it corresponds to a birthrate of 
\begin{equation}
    \delta r=S \Phi(\ln M_1) \varphi(\ln M_2) \Psi(\ln a_2) \delta \ln M_1 \delta \ln M_2 \delta \ln a  \ ,
\end{equation}
where $S$ is the star formation rate assumed to be a constant $S=7.6085 \mathrm{yr}^{-1}$. And the contribution to the expected number of the concerned systems is 
\begin{equation}
    \delta n= \delta r \delta t,
\end{equation}
where $\delta t$ is the lifetime of the concerned systems for each grid.

The primary mass distribution $\Phi(\ln M_1)$ can be derived by 
\begin{equation}
    \Phi(\ln M_1)=M_1\xi(M_1),
\end{equation}
where the $\xi(M_1)$ is the initial mass function(IMF)  given by \cite{kroupa1993distribution},

\begin{equation}
    \xi(m)= \begin{cases}0.29056 m^{-1.3} & 0.1<m \leq 0.5 \\ 0.15571 m^{-2.2} & 0.5<m \leq 1.0 \\ 0.15571 m^{-2.7} & 1.0<m<\infty\end{cases}.
\end{equation}

The distribution of the secondary mass can be obtained from the mass ratio $q_2$,
\begin{equation}
    \varphi\left(\ln M_2\right)=\frac{M_2}{M_1}=q_2,
\end{equation}
which is assumed to satisfy a uniform distribution in the range of $[0,1]$.

The distribution of separation $a$ is assumed to be 
\begin{equation}
    \Psi(\ln a)=0.12328,
\end{equation}
which is normalized within the range of 3 and $10^4 \mathrm{R}_{\sun}$.

In the BSE code, the kick velocity is taken from Maxwellian distribution, and the direction is uniformly distributed in the full solid angles. In previous studies using BSE, each grid point was sampled only once, i.e., only a particular kick velocity direction was simulated. This results in  the problem of under-sampling, because the binary parameters after kick strongly depend on the direction of the kick velocity. Therefore, the outcome of the BPS using BSE change significantly with the random seed for the black holes(BHs) and NS systems which have undergone the kick event\citep{hurley2002evolution,chen2011can}.

To improve this under-sampling problem, we also include the kick direction parameter in the grid of the initial binary parameters. In such a case, the new contribution to the expected number is read,
\begin{equation}\label{eq:expected number}
    \delta n_{kick}=\delta n P(\sin\phi)Q(\omega) \delta\sin\phi\delta\omega,
\end{equation}
where the kick direction is determined by the two angles $\phi$ and $\omega$, defined in the \cite{hurley2002evolution}. Assuming a uniform distribution of  direction in the full solid angles, such an angle distribution is read, 
\begin{equation}
    P(\sin\phi)=0.5
\end{equation}
for $\sin\phi$ in the $[-1,1]$, and 
\begin{equation}
    Q(\omega)=\frac{1}{2\pi}
\end{equation}
for $\omega$ in the $[0, 2\pi]$. The number of the grid points of the two angels of the kick velocity is taken as 10.

The introduction of the kick parameters increases the number of grid points, so that  longer computation time is required. Here we use a two-step approach to perform simulations. Firstly, the simulations of the grid points without kick parameters are made. Secondly, the points where AIC can occur are selected. And finally, each selected grid point is simulated again with a different kick direction. 

Two values of the $\alpha$ are adopted in the CE evolution, $\alpha=1$ and $\alpha=3$. Although the small kick velocity of the WD's AIC is widely adopted in the previous studies \cite{hurley2010formation,tauris2013evolution,freire2013direct}, there are studies adopt large kick velocity \citep[and the references therein]{chen2011can}. Here we adopt both small and large kick velocities, which are chosen as 0, 10, 30, 50, 70, 90, 110, 130, 150, 170, and 190 km/s. The metallicity is taken as $Z=0.02$ and the other parameters in the BSE are the same as those of  \cite{hurley2010formation}.We evolve these binaries for 12Gyr.

\section{Results}
\label{result}
\subsection{ The expected number of eMSPs}

The orbital periods of observed MSPs with helium WD companion are within 0.1 to 1000 days \citep{ginzburg2022eccentric}. Like \cite{hurley2010formation}, we define the detectable systems as those with $0.1<P_b<1000$ days.

\begin{figure}
    \includegraphics[width=\columnwidth]{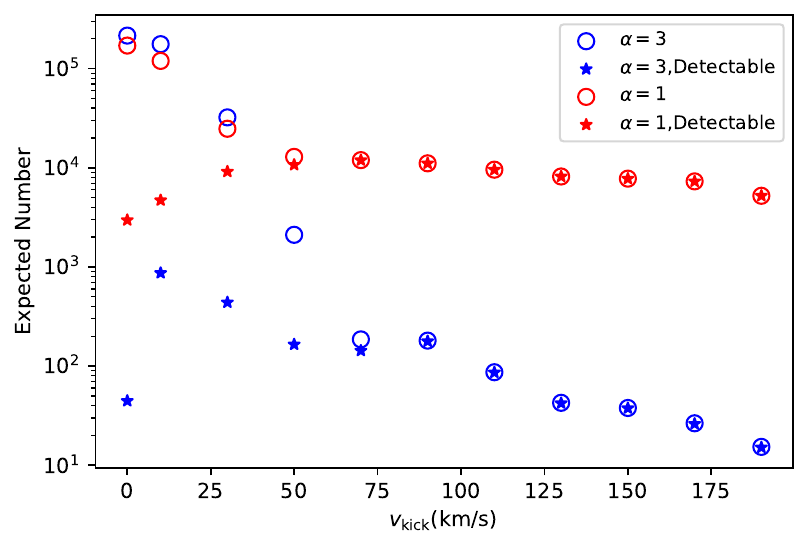}
    \caption{The relation between the kick velocity and expected number of the eccentric MSP-WD systems.\label{fig: relation}}
\end{figure}

The expected number of eccentric MSP-WD systems with each kick velocity is shown in Fig.\ref{fig: relation}. It decreases as the kick velocity increases in both cases of $\alpha=1$ and $\alpha=3$. For the detectable systems, it increases and subsequently decreases as the kick velocity increases. When the kick velocity is small, there are more systems with $\alpha=3$ than those with $\alpha=1$, but for detectable eMSPs, more systems reside in $\alpha=1$ rather than within $\alpha=3$. As the kick velocity increases, the number of systems with $\alpha=1$ and $\alpha=3$ as well as the number of corresponding detectable systems converges, i.e., the undetectable systems disappear. And the number of systems with $\alpha=3$ is much smaller than the number of systems with $\alpha=1$ when the kick velocity is large.

Such a result is a bit counterintuitive. A larger kick speed will lead to a larger eccentricity after the kick event, which in turn will generate more eMSPs\citep{freire2013direct,han2021asymmetrical}. However, this is opposite to the result shown in Fig.\ref{fig: relation}. The reason is that the larger kick velocity will also make the binary easier to be disrupted or become tighter so that it is easier to merge or be circularized.

\subsection{The distribution of eMSPs parameters}
\begin{figure*}
    \includegraphics[width=\linewidth]{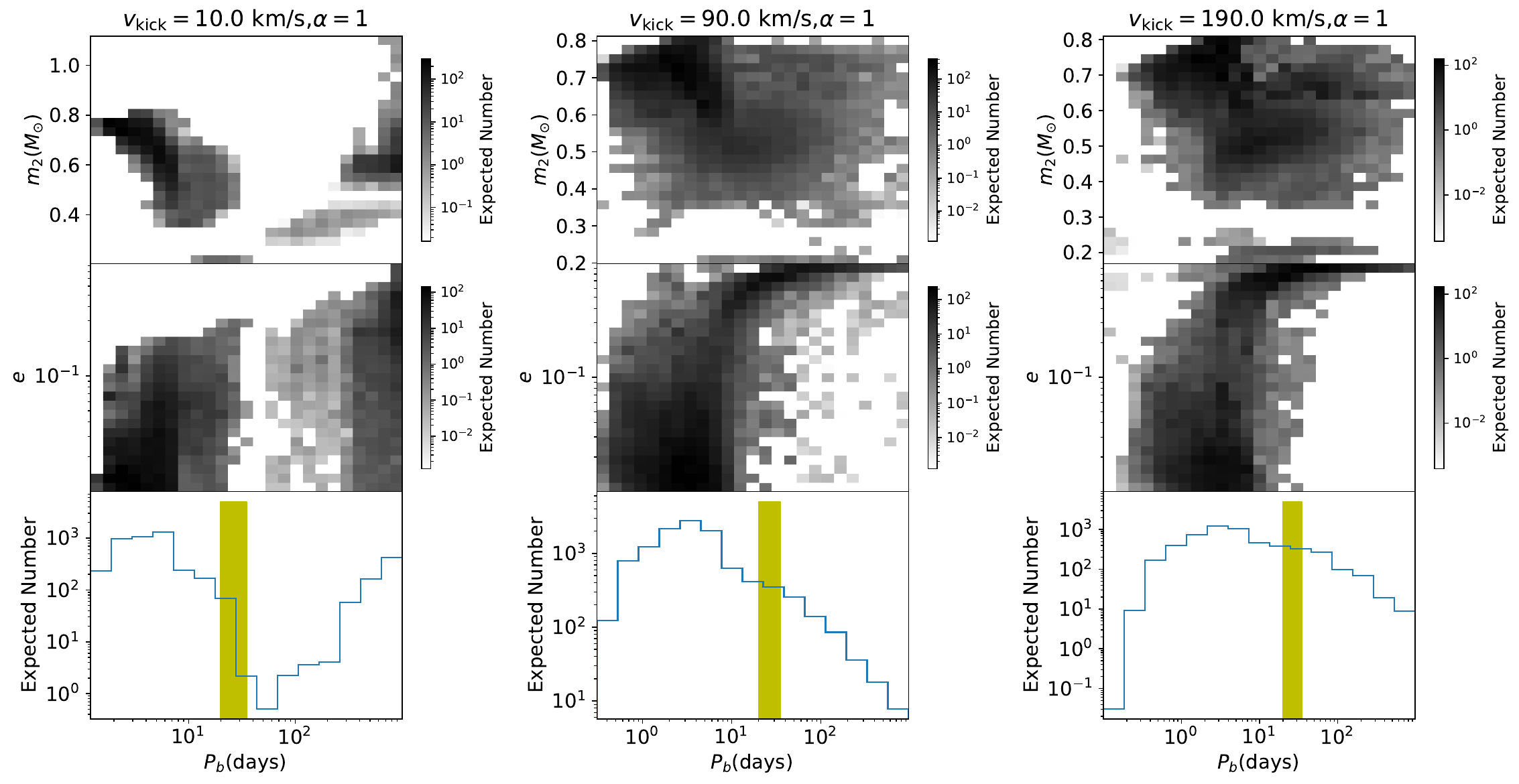}
    \caption{The distribution of the parameters of the detectable eMSPs with $\alpha=1$. The top and middle panels show the distribution of $m_2-P_b$ and $e-P_b$ respectively, and the lower panel shows the distribution of $P_b$. The colorbar represent the expected number defined by Eq.\ref{eq:expected number}. The yellow region represent the orbital period range of 20~35 days, which the observational eMSP locate in. The plots from left to right represent the results with $v_{kick}=$ 10, 90, and 190 km/s respectively.\label{fig: alpha1}}
\end{figure*}

\begin{figure*}
    \includegraphics[width=\linewidth]{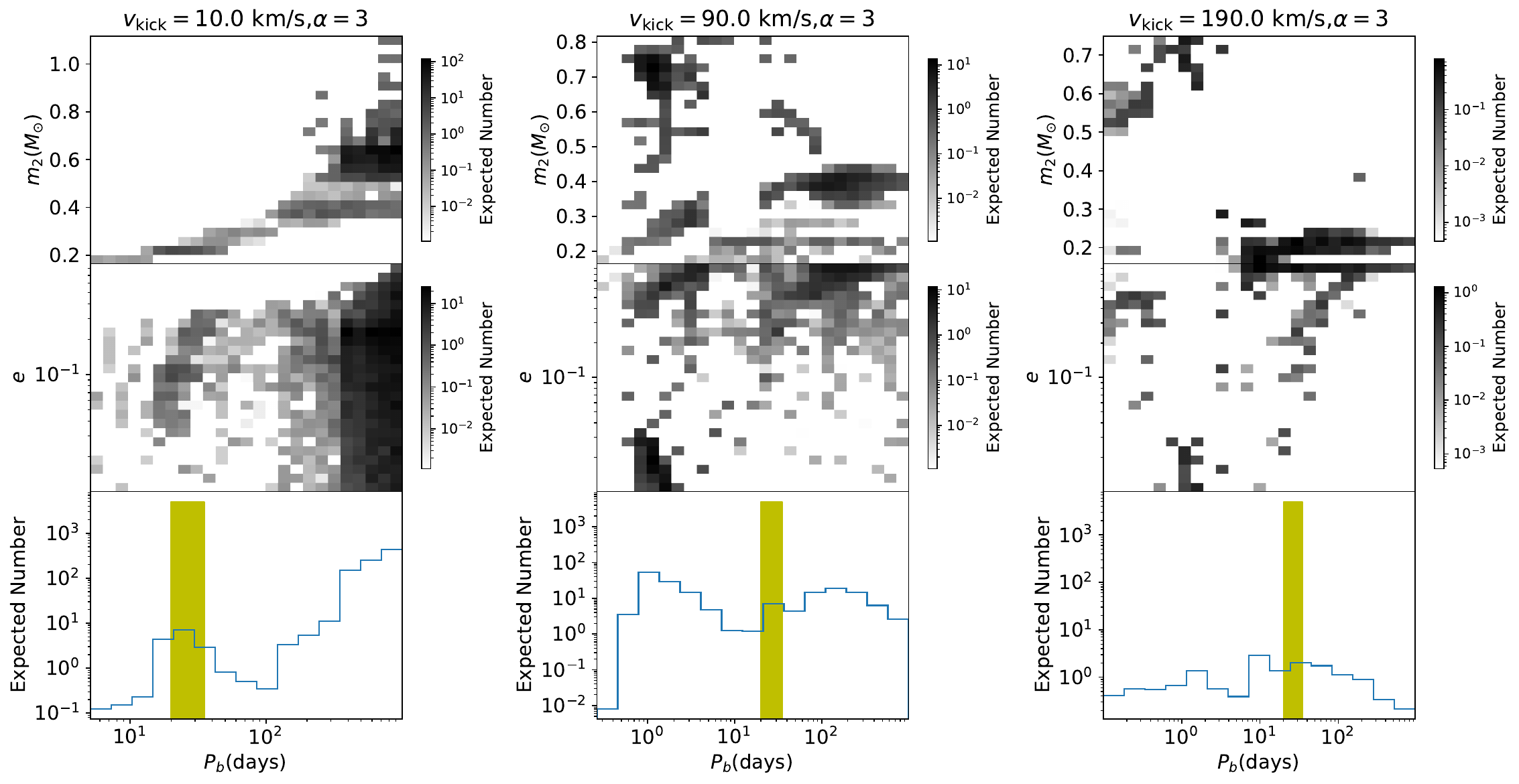}
    \caption{As in Fig.\ref{fig: alpha1}, but with $\alpha=3$.\label{fig: alpha3}}
\end{figure*}

The distribution of $e/m_2-P_b$ of the detectable eMSPs is shown in Fig.\ref{fig: alpha1} and Fig.\ref{fig: alpha3}, corresponding to the cases of $\alpha=1$ and $\alpha=3$ respectively. Such distributions of all eMSPs is shown in the appendix \ref{sec:all_figure} We only exhibit results for $v_{kick}=$ 10, 50 and 190 km/s corresponding to low, medium, and high kick velocity, respectively. There is a significant discrepancy between the results for $\alpha=1$ and $\alpha=3$. The case of  $\alpha=1$ corresponds to the more diffuse distribution of parameters and $\alpha=3$ demonstrates some concentrated areas of parameters. The distribution of parameters for medium and high kick velocities is very similar but differs profoundly from that of low kick velocity. In the case of low kick velocity, the distribution of $P_b$ shows a double-peaked structure for both $\alpha=1$ and $\alpha=3$.

\begin{figure}
    \includegraphics[width=\linewidth]{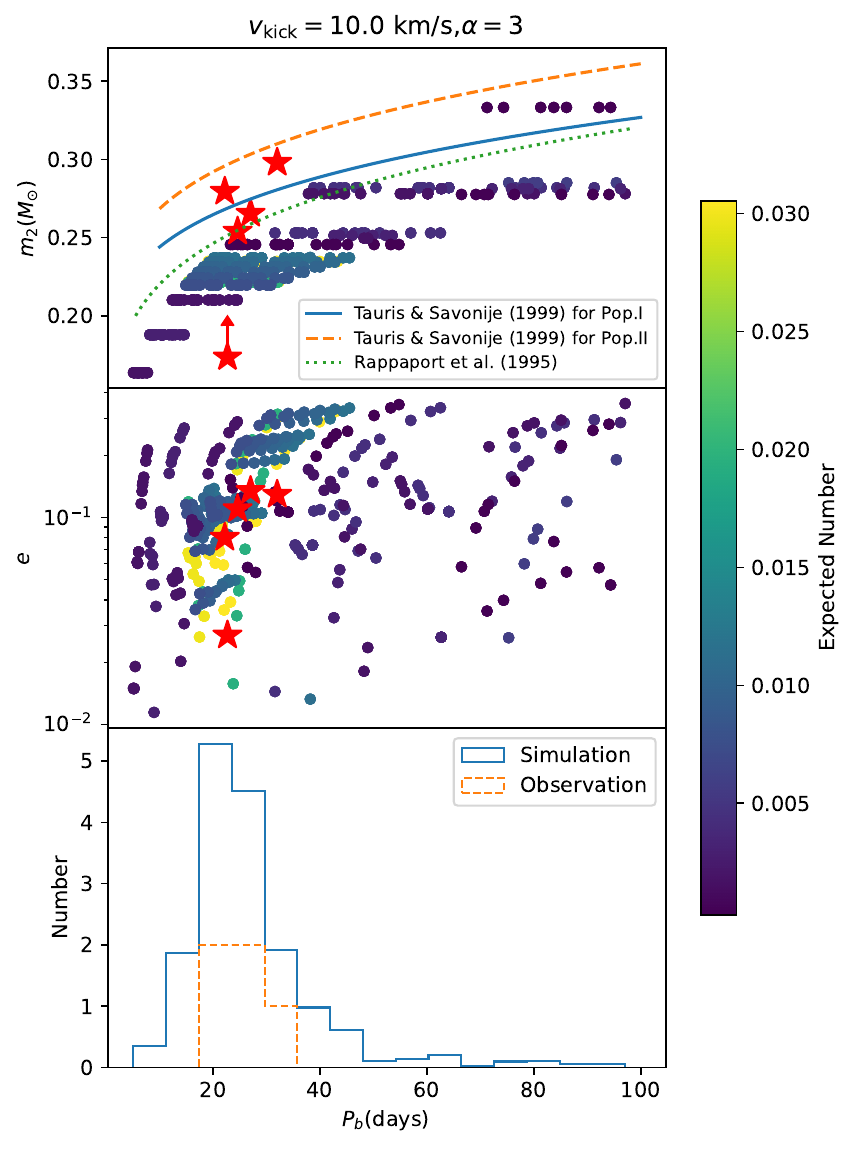}
    \caption{As in Fig.\ref{fig: alpha3} with $v_{kick}=$ 10 km/s, but as a scatter plot to show each simulated point. The solid red star points represent the parameters of the five observed eMSPs, which are taken from \protect\cite{serylak2022eccentric} and the references therein. For the companion mass of PSR J1618-3921, the minimum mass estimated from the mass function is adopted and taken from ATNF catalog \protect\citep{manchester2005australia}. In the top panel, the green, orange, and blue lines represent the $m_2-P_b$ relationships by \protect\cite{rappaport1995relation} and \protect\cite{tauris1999formation} for Pop.I and Pop.II respectively. In the bottom panel, the dashed orange and solid blue lines represent the distribution of the simulated and observed samples respectively. \label{fig: best}}
\end{figure}

The orbital period range of observational eMSP is 20~35 days. The orbital period distribution of the simulation that fits the observation should have a peak near this range. The Fig.\ref{fig: alpha1} and Fig.\ref{fig: alpha3} show that the parameters that have the nearest peak to the observational range are $\alpha=3$ and $v_{kick}=$ 10 km/s. In this case, although the orbital periods of eMSPs are mostly distributed over 100 days, there is a peak below 100 days that appears to match the observations. To compare more intuitively the simulation results corresponding to this peak with the observed eMSPs, the distribution of $e/m_2-P_b$ of the simulated eMSPs with $0.1<P_b<100$ days are shown as a scatter plot in Fig.\ref{fig: best}. With $\alpha=3$ and $v_{kick}=$ 10 km/s, the orbital period distributions of the eMSPs obtained from the simulations well match with those of the observed eMSPs.

However, the simulated companion mass of the eMSPs is smaller than the observed one. This difference essentially represents the deviation of the simulated $m_2-P_b$ relationship obtained from the simulations and the observed values. This deviation may be caused by the kick increasing the orbital period, which also leads to the dispersion of the orbital period in Fig.\ref{fig: best}. But such an effect can also reduce the orbital period, the theoretical curve should cross the distribution of the orbital period. Thus it can not explain this deviation.

The $m_2-P_b$ relationship was first discovered by \cite{rappaport1995relation} and subsequently obtained by \cite{tauris1999formation} with an updated stellar evolution code. Their results are compared with observation and our simulations as shown in Fig.\ref{fig: best}. The $m_2-P_b$ relationship from \cite{tauris1999formation} for Population I ($Z=0.02$) consist best with the observation. The discrepancy between their results stems from the different stellar evolutionary codes they use \citep{tauris1999formation}. Therefore, the difference between the results of ours and theirs may also originate in the same reason.

\begin{figure}
    \includegraphics[width=\linewidth]{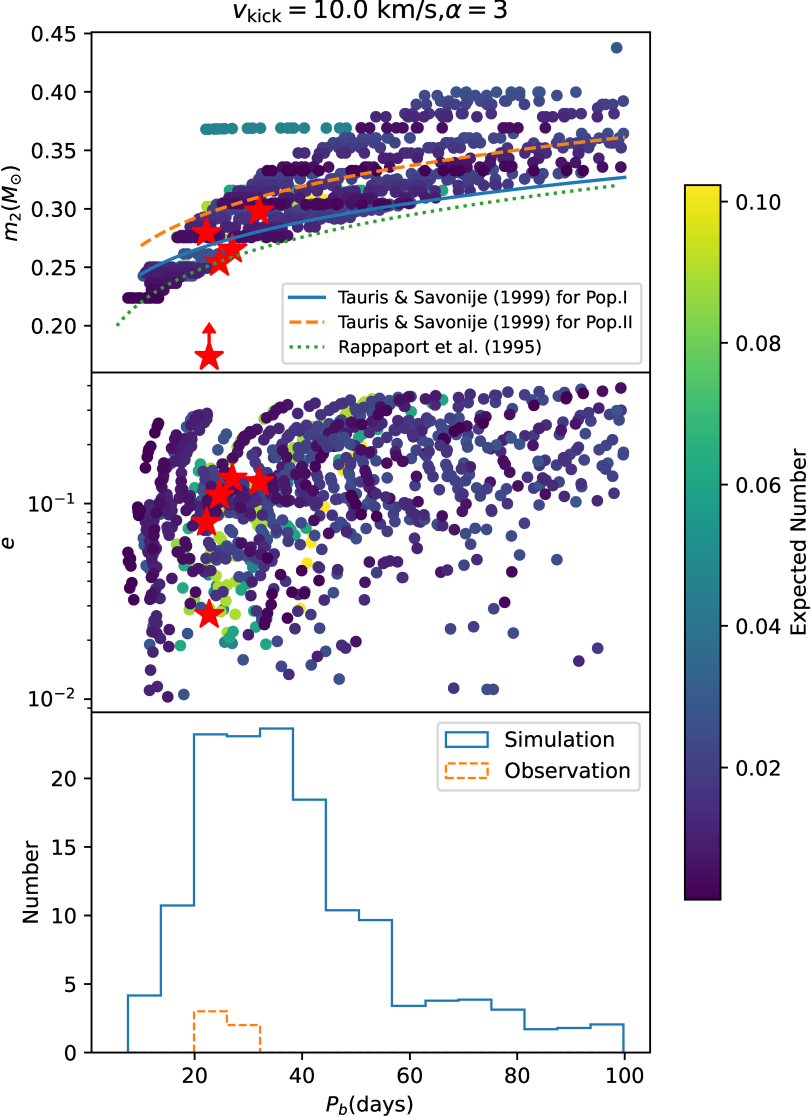}
    \caption{As in Fig.\ref{fig: best}, but for Pop.II($Z=0.001$).\label{fig: best2}}
\end{figure}

The $m_2-P_b$ relationship also depends on metallicity \citep{ergma1998evolution,tauris1999formation}. For a given $P_b$, $m_2$ increases as metallicity decreases. Therefore, simulation with $\alpha=3$ and $v_{kick}=$ 10 km/s for the Population II ($Z=0.001$) are performed as shown in Fig.\ref{fig: best2}. In such a  case, the $m_2-P_b$ relationship obtained from the simulation well agrees with the observed values\footnote{Some points that do not follow the relation satisfied by other points. They correspond to the situation that the companion evolve into the He star and then into the He giant star and finally into the CO WD after the AIC event.}. However, the peaks in the $P_b$ distribution are wider than that with $z = 0.02$, indicating a more diffused  $P_b$ than the observed values. In addition, the number of eMSPs is also greater than those with $z = 0.02$.

Although the simulated $m_2-P_b$ distribution for Pop.II is more consistent with the observations, the $m_2-P_b$ relationship from \cite{tauris1999formation}  suggests that the observations are more consistent with Pop.I. Since \cite{tauris1999formation} uses an updated numerical stellar evolution code, whereas BSE uses a semi-analytic stellar evolution\citep{hurley2002evolution}, the results by \cite{tauris1999formation} should be somewhat more accurate. The observed eMSPs are more likely to have evolved from Pop.II.
\section{Discussion}
\label{discussion}
The result that best fits the observation implies that there are eMSPs with orbital periods between 100 and 1000 days. Considering results for the full orbital period ranges shown in Fig.\ref{fig: alpha3_all}, there are more eMSPs with larger orbital periods. These wide binaries have not been found in the observation. Such systems also appear in the other AIC studies including circular binaries\citep{hurley2010formation,wang2022formation}. The possible reason for such a contradiction is due to the assumption that the NS born through WD's AIC is the MSP. For a NS to be a pulsar, in addition to its radiation cone pointing towards the Earth, its magnetic field $B$ and rotation period $P$ are required to satisfy $ B/P^2>0.17\times 10^{12}\mathrm{G}/\mathrm{s}^2$, which is above the death line \citep{bhattacharya1992decay}. A NS of radius 10km produced by the collapse of a typical WD with radius 3000 km and magnetic field $10^3 \mathrm{G}$ corresponds to a surface magnetic field of $10^8 \mathrm{G}$ \citep{freire2013direct}, which is the typical value for MSP.

The evolution of the NS's rotation period is ignored in our simulation, which depends on two factors: one is the initial rotation period determined by the rotation of the WD as the progenitor, and the other is the NS's rotation evolution during NS accretion\citep{chen2011can}. Since the eMSP discussed in this paper is insufficiently circularized, which implies that the lifetime of the NS during the accretion phase is short, the latter can be neglected. For the former, the rotation evolution of the WD needs to be considered, which may affect the mass of the NS due to the possible differential radiation of the WD \citep{yoon2004presupernova,yoon2005evolution}.

Two of the observed eMSPs have pulsar mass of more than $1.44 \mathrm{M}_{\sun}$ \citep{zhu2019mass,barr2017massive}, which is the maximum mass of WDs in our simulations. Since the mass of the eMSPs produced by AIC is smaller than those of the progenitors, our simulations cannot produce MSPs with such a large mass. Even in the RD-AIC model, the white dwarf can only have a maximum mass of $1.48 \mathrm{M}_{\sun}$, which is the maximum mass of a rigidly rotating WD \citep{freire2013direct}. However, if the white dwarf has a differential rotation, it can have a maximum mass of up to $4 \mathrm{M}_{\sun}$ \citep{yoon2004presupernova}. The angular momentum carried by the WD accreting material is first transferred to the outer layers of the WD, and then to the inner sides, thus redistributing the angular momentum. If the transfer of angular momentum is not efficient enough, the WD in accretion will be in a differential rotation rather than a rigid one.

The evolution of the differential accreting WD has been studied by \cite{yoon2005evolution}. The evolutionary path of the accreting WD depends on the angular momentum accretion efficiency, the angular momentum transport efficiency, and the angular momentum loss due to gravitational radiation induced by the r-mode instability. Whether accretion terminates before the WD reaches the critical mass will determine whether the collapse is delayed. The RD-AIC model is the delayed collapse of the rigidly rotational WD, which is one of the cases that \cite{freire2013direct,yoon2005evolution} considered.

The initial rotation period of the NS is also likely to be constrained by the observation. The AIC of the WD is one of the possible origins of the fast blue optical transients(FBOT) \citep{yu2019optical,yu2019x}. And \cite{liu2022magnetar} found that the rotation period of the remnant NS after FBOT event is $9.1_{-4.4}^{+9.3} \mathrm{~ms}$, which slightly matches the observational value of the eMSPs. However, the magnetic field of the NS they found is very large $11_{-7}^{+18} \times 10^{14} \mathrm{G}$, which contradicts our assumptions.

The change of mass limit of the WD with differential rotation will significantly the evolution of the AIC or SNe Ia, making the parameter space smaller\citep{chen2009progenitors}. Although our results well explain the narrow orbital period range of the observation, it will make a difference if the rotation evolution is considered. Even so, due to the required short lifetime of the companion after the kick event, the results of the simulations with the rotation evolution are expected to be able to have a narrow orbital period distribution for some specific parameters.

The AIC evolution of the WDs contains direct AIC of the Chandrasekhar WDs, the direct and delay AIC of the super-Chandrasekhar WD \citep{yoon2005evolution}. And the rotation of the WDs also affects the rotation of the newborn NS. Thus a comprehensive simulation of the AIC evolution of the WD taking into account the rotation evolution of the WD and the NS is necessary, which will be included in our future simulations.

The direct AIC of the super-Chandrasekhar WD with insufficient circularization is a possible way to form PSR J0955-6150. The pure kick simulation show that the kick event can change the orbital period to the observational value of 24.58 days \citep{serylak2022eccentric}. The larger eccentricity than observation induced by such kick event can decrease to the observational value of 0.11 by tidal circularization while the companion evolves to the white dwarf. However, the circularization will also decrease the orbital period, so whether the final orbital period can match observation is questionable. 

\cite{serylak2022eccentric} also found that the spin axis of the MSP does not align with the direction of the orbital angular momentum vector, and the misalignment angle $\delta >4.8$ deg is surprisingly huge. Such a misalignment angle can also be caused by the kick event which can change the direction of the orbital angular momentum, then constrain the kick velocity.

Here we use method described by \cite{hurley2002evolution} to calculate the orbital period $P_{b,0}$, eccentricity $e_0$ and misalignment angle after kick event. Following the calculation of \cite{serylak2022eccentric}, the mass of the white dwarf and orbital period in the pre-AIC stage is set to 1.96 $\mathrm{M}_{\sun}$ and 14.2 days(Pop. I), respectively. After the kick event, the circularization will decrease both the orbital period and eccentricity. In this process, orbital angular momentum $L=\sqrt{GMa(1-e^2)}$ is roughly conserved, thus the final orbital period can be derived by $P_{b,f}=\left(\frac{1-e_f}{1-e_0}\right)^{\frac{3}{2}}P_{b,0}$, where the final eccentricity $e_f$ is the observational value 0.11.

\begin{figure}
    \includegraphics[width=\linewidth]{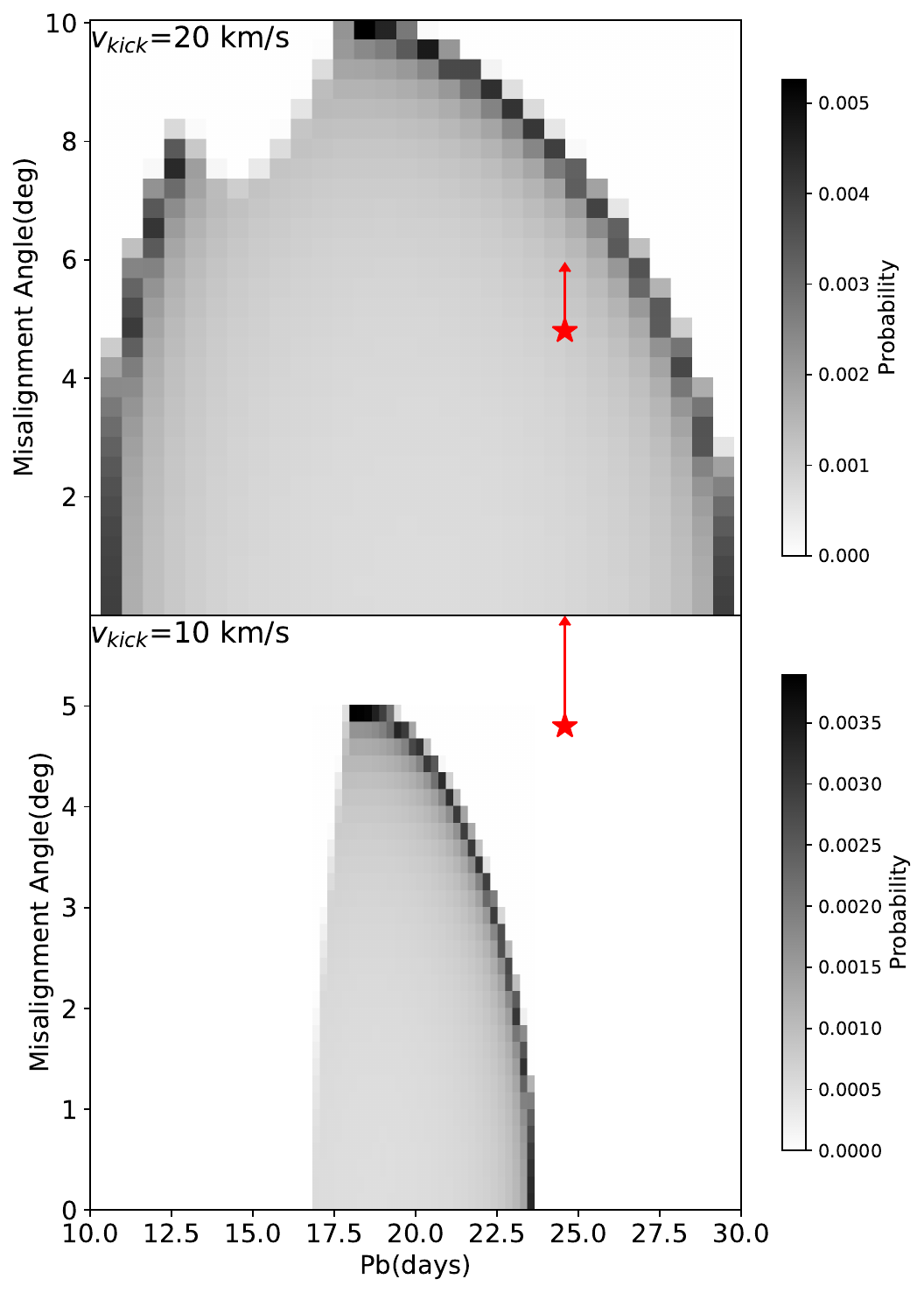}
    \caption{The distribution of misalignment angle and orbital period of the simulation for direct AIC with insufficient circularization process of PSR J0955-6150. The red star represents the parameters of PSR J0955-6150. The probabilities are calculated under the assumption of isotropic kick velocity direction.\label{fig: misalignment}}
\end{figure}

The simulation result is shown in Fig.\ref{fig: misalignment}. Only post-AIC eccentricity $e_0>e_f=0.11$ is considered. while kick velocity is 10 km/s, the final orbital period can not match observation even if allow all misalignment angles. For the higher kick velocity 20 km/s, the final orbital period range is broadened, and both the final orbital period and misalignment can match observations. The edges of the parameter space have the maximum probability, thus the kick velocity is likely between 10 and 20 km/s.

\section{Conclusion}
\label{summary}

We suggest that the eMSP is formed from the AIC of WD under insufficient circularization. By performing a binary population synthesis with BSE, we find that the distribution of $e$ and $m_2$ with $\alpha=3$ and $v_{\mathrm{kick}}=10$ km/s can reproduce the observational peak in the distribution of the orbital period. In particular, it allows a natural interpretation of the narrow range of orbital periods of the observational eMSPs. In addition, we find that smaller kick velocities are more likely to produce eMSPs, which is somewhat counterintuitive.

Our results also show that there are many binaries with large orbital periods, which are not found in the observation. This issue may arise from our artificial assumption that all NSs formed by AIC are MSPs. Some of them possibly do not rotate fast enough to emit radiation, which requires the rotation evolution of the WD and MSP to be taken into account in the simulation.

Despite our simulation ignoring them, our results still show that the small parameter space corresponding to insufficient circularization can reproduce the narrow orbital period range in the observation. Such a result indicates that this scenario is a potential mechanism to form observational MSPs. More realistic simulations in the future need to include direct and delay AIC of the super-Chandrasekhar WD, which is expected to solve the existing problems of wide binaries and large pulsar mass.

\section*{Acknowledgements}
This work was supported by the National Key Research and Development Program of China(No. 2020YFC2201400) and National SKA Program of China (2020SKA0120300). We acknowledge Beijng PARATERA Tech CO.,Ltd. for providing HPC resources that have contributed to the research results reported within this paper.

\section*{Data Availability}
The result and code of this work are available from the corresponding authors at reasonable request.
\bibliographystyle{mnras}
\bibliography{ref} 
\appendix
\section{ Results for the full orbital period range}
\label{sec:all_figure}
\begin{figure*}
    \includegraphics[width=\linewidth]{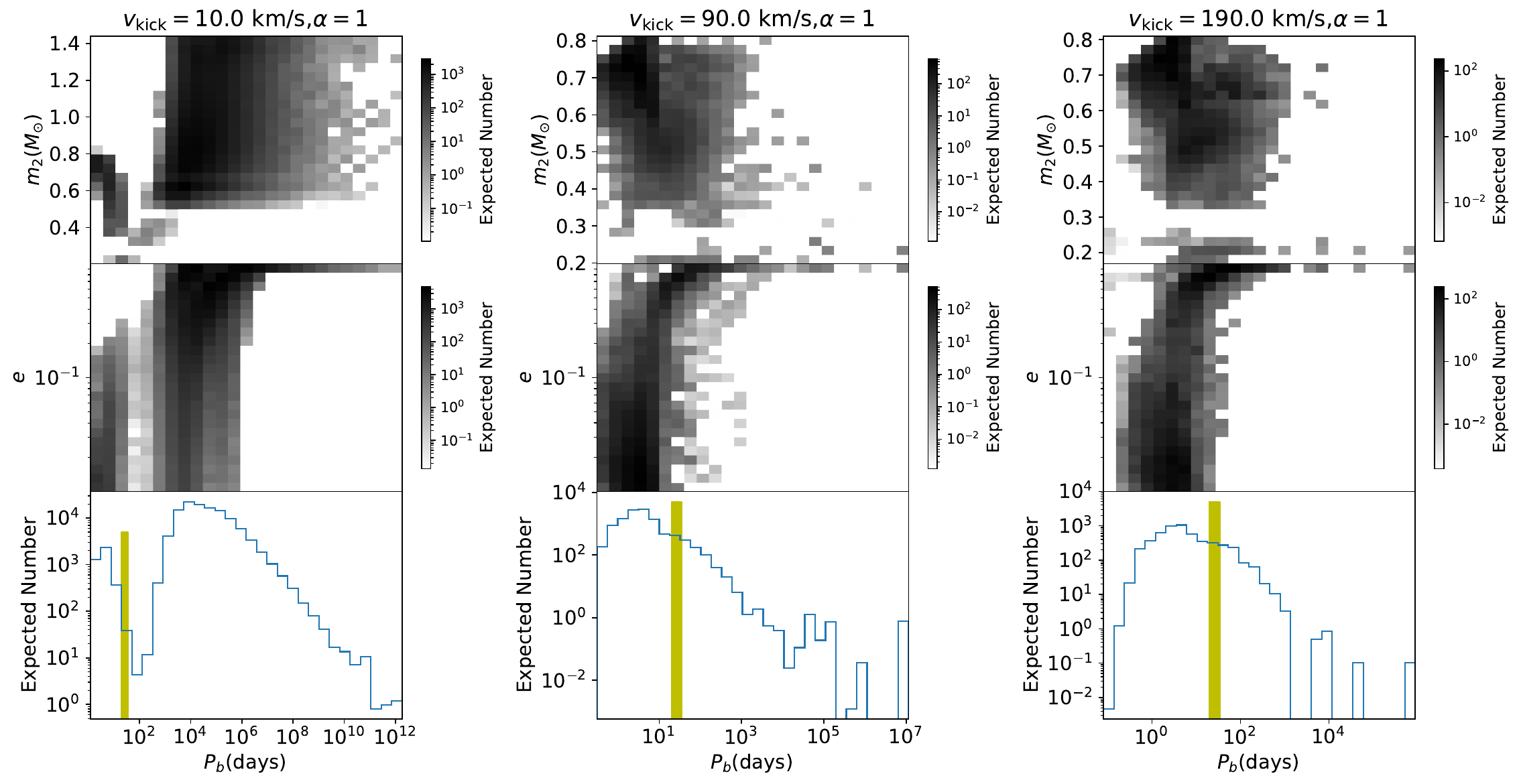}
    \caption{As in Fig.\ref{fig: alpha1}, but for the full orbital period range.\label{fig: alpha1_all}}
\end{figure*}
\begin{figure*}
    \includegraphics[width=\linewidth]{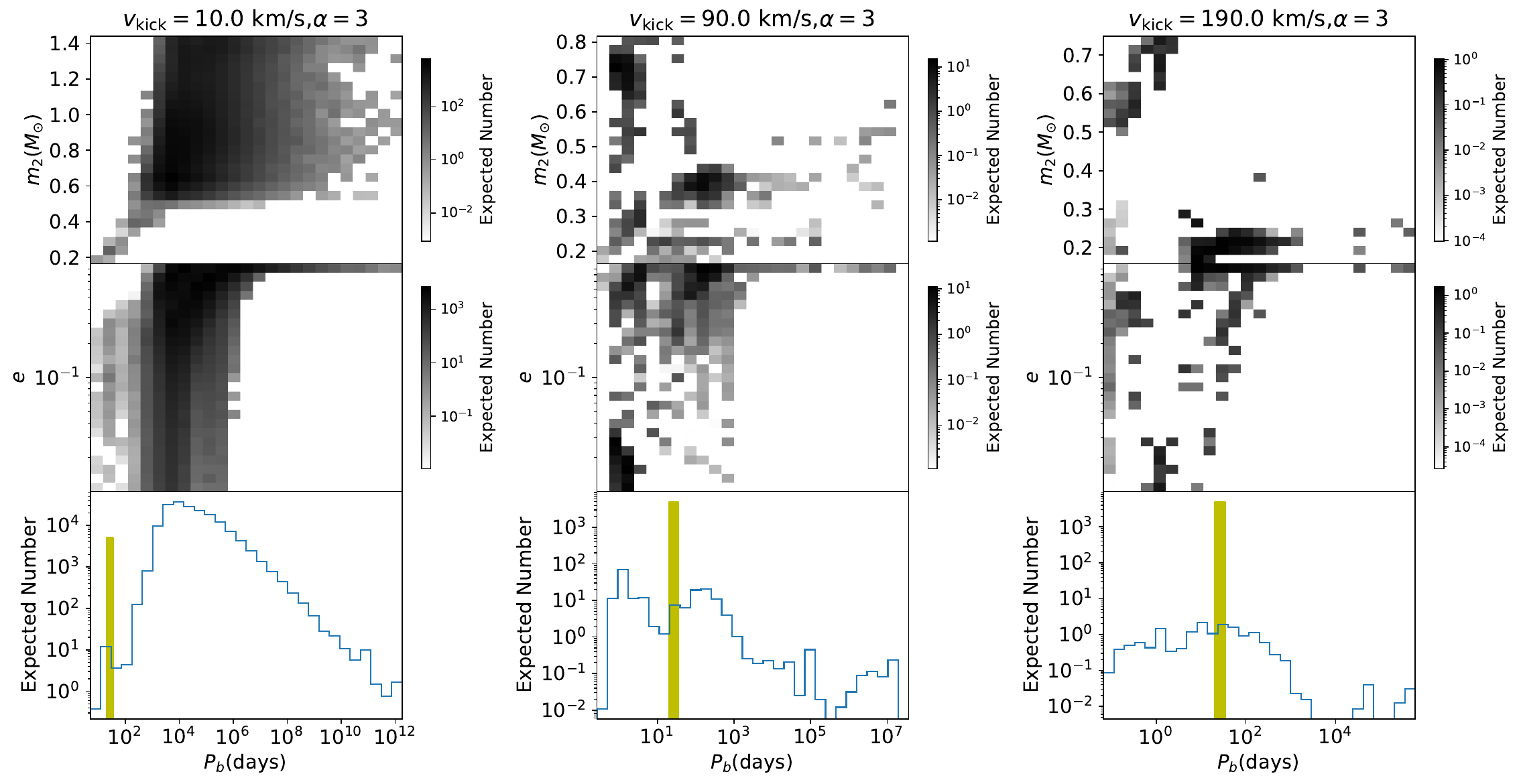}
    \caption{As in Fig.\ref{fig: alpha1_all}, but with $\alpha=3$.\label{fig: alpha3_all}}
\end{figure*}
\end{document}